\documentclass[letterpaper,conference]{IEEEtran}

\IEEEoverridecommandlockouts
\usepackage{graphicx}
\usepackage{epstopdf}
\usepackage[caption=false,font=footnotesize]{subfig}
\usepackage{cite}
\usepackage{color}
\usepackage[cmex10]{amsmath}
\usepackage{algorithm}
\usepackage{algpseudocode}
\usepackage{threeparttable}
\usepackage{multirow}

\newcommand{\mtx}[1]{\ensuremath{\boldsymbol{#1}}}

\begin{document}

\title{Histogram-Based Flash Channel Estimation}
\author{
Haobo Wang, Tsung-Yi Chen, and Richard~D.~Wesel\\
whb12@ucla.edu, tsungyi.chen@northwestern.edu, wesel@ee.ucla.edu
\thanks{ This work was supported by a grant from Western Digital Corporation and by the Center on Development of Emerging Storage Systems (CoDESS) http://www.uclacodess.org.}
}
\maketitle
% ------------ABSTRACT-----------------------
\begin{abstract}
Current generation Flash devices experience significant read-channel degradation from damage to the oxide layer during program and erase operations.   Information about the read-channel degradation drives advanced signal processing methods in Flash to mitigate its effect.  In this context, channel estimation must be ongoing since  channel degradation evolves over time and as a function of the number of program/erase (P/E) cycles.  This paper proposes a framework for ongoing model-based channel estimation using limited channel measurements (reads).  This paper uses a channel model characterizing degradation resulting from retention time and the amount of charge programmed and erased.   For channel histogram measurements, bin selection to achieve approximately equal-probability bins yields a good approximation to the original distribution using only ten bins (i.e. nine reads).  With the channel model and binning strategy in place, this paper explores candidate numerical least squares algorithms and ultimately demonstrates the effectiveness of the Levenberg-Marquardt algorithm which provides both speed and accuracy.
\end{abstract}
% ------------END ABSTRACT--------------------
% ------------INTRODUCTION-----------------------
\section{Introduction}
\label{Section:Introduction}
With widespread use in computers, phones and even satellites, Flash memory has become one of the key components directly contributing to the fast-paced evolution of electronic systems. However, the reliability of Flash memory degrades with respect to usage and retention time. Both the extent of usage and the retention time during which data can be reliably recovered are limited. This causes significant drawbacks in practical applications. Furthermore, physical cell density and modulation constellation density are increasing rapidly to satisfy the demand for increased storage capacity under strict physical size constraints.  This amplifies the degradation problem.

Modern Flash storage solutions employ channel codes \cite{maeda_error_2009,klove_systematic_2011} to increase reliability, but this approach alone cannot effectively counteract the channel capacity decrease caused by degradation due to recursively program and erase the cells.  Signal processing methods such as Dynamic Voltage Allocation (DVA)\cite{chen_increasing_2014} and Dynamic Threshold Assignment (DTA) \cite{sala_dynamic_2013} actively mitigate channel degradation.  Effective decoding of channel codes, possible selection of channel code rate, and adaptive signal processing such as DVA and DTA require channel state information.  Thus, it is necessary to have a robust on-line estimation framework for this evolving channel.

\begin{figure}[!t]
\centering
\includegraphics[width=0.4\textwidth]{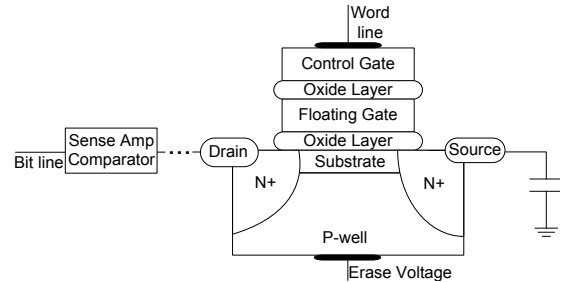}
\caption {Common structure of a P-well Flash memory cell.}
\label{Figure:FlashCell}
\vspace{-3ex}
\end{figure}

Figure \ref{Figure:FlashCell} shows basic physical structure for a P-well Flash memory cell. The threshold voltage of a Flash cell depends on the amount of charge in the floating gate and the device's intrinsic threshold voltage when the floating gate is empty. Electrical charge can be programmed to and erased from the floating gate through the oxide layer. As a result, the threshold voltage can be controlled by program and erase (P/E) operations \cite{bez_introduction_2003}.  Information about the threshold voltage can be obtained by applying a word line voltage and reading the sense amplifier output to determine if the threshold voltage has been surpassed.  Due to channel degradation, the measured value of the threshold voltage often differs from the originally stored threshold voltage.  

This paper explores histogram-based channel estimation (HBCE) for Flash.  For a given block, multiple sense amplifier reads at distinct wordline voltages yield a histogram of the operational threshold voltages. This histogram can provide an estimate for the actual read channel distribution. 

In \cite{lee_estimation_2013}, the authors demonstrate that a multimodal Gaussian distribution representing four-level MLC Flash can be accurately estimated by least squares (LS) with 12-bin histograms using the Levenberg-Marquardt (LM) algorithm. In \cite{lee_estimation_2013} eight parameters are estimated, four means and four variances. 

Our work is largely inspired by \cite{lee_estimation_2013}.   We apply HBCE to a read channel distribution that models the device physics of Flash degradation, namely  wear-out and retention-loss.  By modeling the underlying physical properties, our channel model uses fewer parameters (five) to characterize the channel.  The combination of fewer parameters and the use of an improved binning paradigm allows LM to accurately estimate a more detailed channel distribution using only 10-bins.

The remainder of this paper is organized as follows: Section \ref{Section:FlashMemoryChannelModel} presents the Flash channel model used in the paper. Section \ref{Section:BinningStrategy} compares three binning strategies from the perspectives of squared Euclidean distance and effective resolution.  Section \ref{Section:ChannelParemeterEstimation} compares three least squares channel parameter estimation algorithms in terms of their speed and accuracy in estimating channel parameters that minimize the squared Euclidean distance between the measured histogram and the histogram produced by the estimated channel parameters in the context of our channel model. Section \ref{Section:ExperimentalResults} presents simulation results demonstrating that ten bins (nine reads) using an equal-probability binning strategy and the Levenberg-Marquardt least-squares algorithm provides excellent channel parameter estimation. Section \ref{Section:Conclusion} concludes the paper..

% ------------END INTRODUCTION-----------------------
\section{Flash Memory Channel Model}
\label{Section:FlashMemoryChannelModel}

Recent research in Flash provides many good channel models, we use the model presented in \cite{chen_increasing_2014} as the basis of our analysis. 

\subsection{Degradation Mechanism}
\label{Section:FlashMemoryChannelModel:DegradationMechanism}
Based on the literature \cite{lee_data_2003, monzio_compagnoni_random_2009, mielke_recovery_2006}, two important forms of degradation are investigated in our analysis. The first mechanism is called wear-out, which is the P/E cycling-related degradation. The second mechanism, which is called retention loss, is also caused by the P/E cycling.  The key distinction between wear-out and retention loss is that variations in threshold due to wear-out occur immediately after writing and variations in threshold due to retention loss occur over the retention time becoming more severe with longer retention times.  

Because Flash cells are densely packed in a two dimensional array in the chip, the coupling effect among the cells and the share of electrical connections cause distortion of the channel known as cell-to-cell interference. This interference depends on the specific structure of individual Flash chip and the operation sequence of the controller\cite{cernea_34_2009,cai_program_2013}. As a result, generalization of our channel model to include such disturbance is highly implementation dependent. Thus, in this paper, cell-to-cell interference is not modeled.  Many methods counteracting this problem have been proposed in the semiconductor community, such as P/E sequence optimization and write voltage pre-distortion\cite{cai_program_2013,dong_using_2013}.  

Similarly, read disturb is implementation dependent and not included in our model.  However, given the implementation details, we believe that both read disturb and cell-to-cell interference could be incorporated in a general channel model.

\subsection{Channel Model}
\label{Section:FlashMemoryChannelModel:ChannelModel}
Based on the theoretical analysis and experimental results from the literature \cite{takeuchi_double-level-vth_1996,compagnoni_first_2007,dong_using_2013,lee_data_2003,monzio_compagnoni_random_2009,mielke_recovery_2006}, our Flash memory channel model is formulated with three additive noise components:
\begin{equation}
\label{Equation:AdditiveModel}
y=x+n_p+n_w+n_r \, ,
\end{equation}
where $x$ is the intended threshold voltage written to a cell, and $y$ is the measured threshold voltage. The noise $n_p$ is the programing noise component related only to the P/E process, $n_w$ denotes the wear-out noise caused by wear-out effect, and $n_r$ represents the retention noise caused by retention loss.

\begin{figure}
\centering
\includegraphics[width=0.35\textwidth]{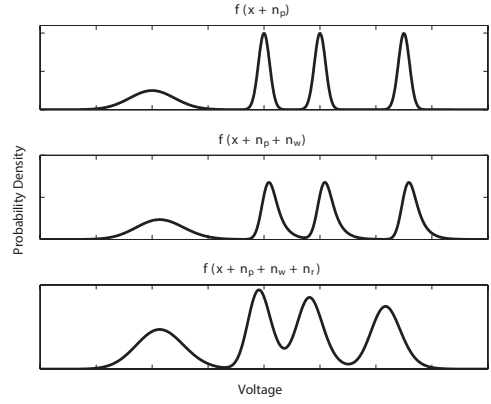}
\caption {Flash read channel probability density functions illustrating degradation mechanisms.}
\label{Figure:ChannelPDFExample}
\vspace{-3ex}
\end{figure}

Figure \ref{Figure:ChannelPDFExample} shows an example channel distribution, which demonstrates the impact of the noise components to the channel.

\subsubsection{Programming Noise}
\label{Section:FlashMemoryChannelModel:ChannelModel:ProgrammingNoise}
The uncertainty of programmed and erased state threshold voltages immediately after P/E operations of a new cell in current generation Flash memory can be modeled by Gaussian distribution. The variance of the distribution depends on cell's stored state, \cite{takeuchi_double-level-vth_1996,compagnoni_first_2007,dong_using_2013}. Let $l$ denote the level of intended threshold with $l=0$ representing the erased state and $l>0$ representing programmed states. Programming Noise can then be modeled as
\begin{equation}
\label{Equation:ProgrammingNoiseModel}
f_{N_P}(n_p|l)=
\begin{cases}
\mathcal{N}(0,\sigma^2_p) & \text{if } l=0,\\
\mathcal{N}(0,\sigma^2_e) & \text{if } l>0.
\end{cases}\ \ \     
\text{where }\sigma_e>\sigma_p \, .
\end{equation}
The noise variance of the programmed states is significantly smaller that the variance for the erased state because of a tight programming feedback loop \cite{compagnoni_first_2007}. This is modeled by having $\sigma_e>\sigma_p$.
 
\subsubsection{Wear-out Noise}
\label{Section:FlashMemoryChannelModel:ChannelModel:WearoutNoise}
Wear-out noise is caused by recurring P/E operations damaging the oxide layer through generation of oxide traps and interface traps. In \cite{lee_data_2003}, the authors point out that interface traps have the most significant impact on wear-out in deeply scaled devices; therefore, the impact of oxide traps is not considered in this component. 

The wear-out effect of traps on measured thresholds can be modeled as Random Telegraph Noise (RTN). RTN widens the distribution of read thresholds with exponential tails on both sides of the actual threshold voltage \cite{monzio_compagnoni_random_2009}. However, based on the data from real devices, the distribution of read thresholds features significant single sided exponential tails in the positive direction, thus we use the following exponential distribution as the model for wear-out noise:
\begin{equation}
\setcounter{equation}{3}
\label{Equation:WearoutNoiseModel}
f_{N_W}(n_w)=
\begin{cases}
\frac{1}{\lambda}e^{-\frac{n_w}{\lambda}} & \text{if } n_w\geq0\\
0 & \text{if } n_w<0
\end{cases} \, ,
\end{equation}
where $\lambda$ is the channel parameter which functions as a metric for interface trap density.

\subsubsection{Retention Noise}
\label{Section:FlashMemoryChannelModel:ChannelModel:RetentionNoise}
Retention loss is caused by electron detrapping.  Thus, the characteristic of this noise component is determined by the interface trap density, oxide trap density and retention time \cite{mielke_recovery_2006}.   Following \cite{mielke_recovery_2006}, we model retention loss as Gaussian noise with distribution 
\begin{equation}
f_{N_R}(n_r)=\frac{1}{\sigma_r\sqrt{2\pi}}e^{-\frac{(n_r-\mu_r)^2}{2\sigma^2_r}} \, ,
\end{equation}
where
\begin{IEEEeqnarray}{l}
\mu_{r}=\gamma_{\mu_r}(x-x_0) \, ,\\
\sigma_{r}=\gamma_{\sigma_r}\sqrt{x-x_0} \, . 
\end{IEEEeqnarray}
The parameters $\gamma_{\mu_r}$ and $\gamma_{\sigma_r}$ represent trap density and retention time.  Variable $x$ is the intended threshold, and $x_0$ is the intended erased-state threshold.  Note that in this model, there is no retention loss if the intended threshold is that of the erased state.

\subsubsection{Overall Conditional Distribution for Flash Channel}
\label{Section:FlashMemoryChannelModel:ChannelModel:OverallConditional DistributionForFlashChannel}
From the discussion above, the conditional distribution for Flash read channel can be summarized as
\begin{equation}
\label{Equation:ConditionalDistribution}
\setcounter{equation}{7}
f_{Y|X}(y|x)=\frac{e^{\frac{\mu_r+x-y}{\lambda}+{\frac{\sigma^2}{2\lambda^2}}}}{\lambda}\cdot Q\left(\frac{\mu_r+x-y}{\sigma}+\frac{\sigma}{\lambda}\right) \, ,
\end{equation}
where $x$ is the intended threshold voltage and $y$ is the measured threshold voltage.  The value of $\sigma$  depends on the intended voltage as follows:
\begin{equation}
\sigma = 
\begin{cases}
\sqrt{\sigma^2_e+\sigma^2_r} & \text{if $x=x_0$}\\
\sqrt{\sigma^2_p+\sigma^2_r} & \text{otherwise}
\end{cases} \, .
\end{equation}
$Q(m)$ is the tail probability of the standard normal distribution: $Q(m)=\int_{m}^{\infty}{\frac{e^{-x^2/2}}{\sqrt{2\pi}} \,dx}$. The parameters $\sigma_r, \mu_r$ and $\lambda$ evolve over both time and P/E cycling process, as the channel degrades, while $\sigma_e$ and $\sigma_p$ remain constant. Given a specific retention time, P/E cycling condition, and the physics related parameters, the exact value of the current channel parameters can be determined using channel parameter degradation model proposed in \cite{chen_increasing_2014}. 

\section{Binning Strategy for Histogram Measurement}
\label{Section:BinningStrategy}
A good binning strategy, i.e. selecting the proper number and placement of word line voltages for the reads that will create the histogram bins, is critical for the efficiency and accuracy of practical histogram-dependent signal processing methods in Flash. 

\subsection{Number of Bins}
\label{Section:BiningStrategy:NumberOf Bins}
A fairly accurate channel estimation can, of course, be derived from a complete voltage scan which uses the smallest possible bin sizes by reading at every available voltage level using the so-called debug mode.  However, the large number of reads required by this process significantly increases device read latency and stalls normal operations.   Such a large number of reads is likely not necessary.  From the soft decoding literature  \cite{wang_soft_2011,wang_enhanced_2014,lee_estimation_2013}, a relatively small number of word-line voltages sufficiently gives good performance in terms of both decoding and channel estimation.

Furthermore, too many bins in the histogram increase the computational complexity in each iteration of the least square algorithms described in Section \ref{Section:ChannelParemeterEstimation} below, and also require more storage space in the controller. Thus a relatively small number of bins can reduce both complexity and latency.  The choice for the number of bins required  also depends on the channel estimation algorithm employed.  Basic algorithms usually require more detailed channel measurements than advanced algorithms. As shown below in Sec. \ref{Section:ExperimentalResults},  a 10-bin histogram can provide accurate estimation of the channel parameters in our model.  Thus, in exploring the performance of the three bin-placement paradigms, we focus on the 10-bin case.

\subsection{Selecting a Bin-Placement Paradigm}
\label{Section:BinningStrategy:HistogramType}

We will consider three bin-placement paradigms: equal-width, equal-probability, and maximum mutual information (MMI).  Equal-width histograms have bins covering intervals of equal length except for the semi-infinite bins at the edges. Equal-probability histograms allocate bins having the same probability (same number of counts), to the extent that this can be achieved without a-posteriori knowledge.  MMI histograms proposed in \cite{wang_soft_2011,wang_enhanced_2014} optimize decoding performance by maximizing the mutual information between the intended threshold voltages and the bin location identified by the reads.

As presented in Section \ref{Section:ChannelParemeterEstimation}, channel parameters are estimated by minimizing the squared Euclidean distance between the measured histogram acting as the reference and the histogram constructed with the estimated channel parameters. To achieve good estimation accuracy, the measured histogram should be as close to the original channel distribution as possible. To compare bin-placement paradigms, the squared Euclidean distance between the channel distribution $f(y)$ and the histogram induced by $f(y)$ is used as the metric to evaluate the amount of discretization error of each bin-placement paradigm. This metric $D_{E^2}$ is defined as follows:
\begin{equation}
\label{Equation:SquaredEuclideanDistance_PDF&HistogramDensity}
D_{E^2}={\sum_{i=0}^{M-1}{\int_{q_i}^{q_{i+1}}{\left(f(y)-\frac{H_i}{q_{i+1}-q_i}\right)^2dy}}} \, ,
\end{equation}
where $f(y)$ is the true read channel distribution, $M$ is the number of bins, and $q_i, q_{i+1}$ represent the left and right boundary of the the $i$th interval.  $H_i$ is the probability of $i$th bin induced by $f(y)$, $H_i = \int_{q_i}^{q_{i+1}}f(y)dy$, and $\frac{H_i}{q_{i+1}-q_i}$ denotes the probability density of the $i$th bin.

\begin{figure}
\centering
\includegraphics[width=0.35\textwidth]{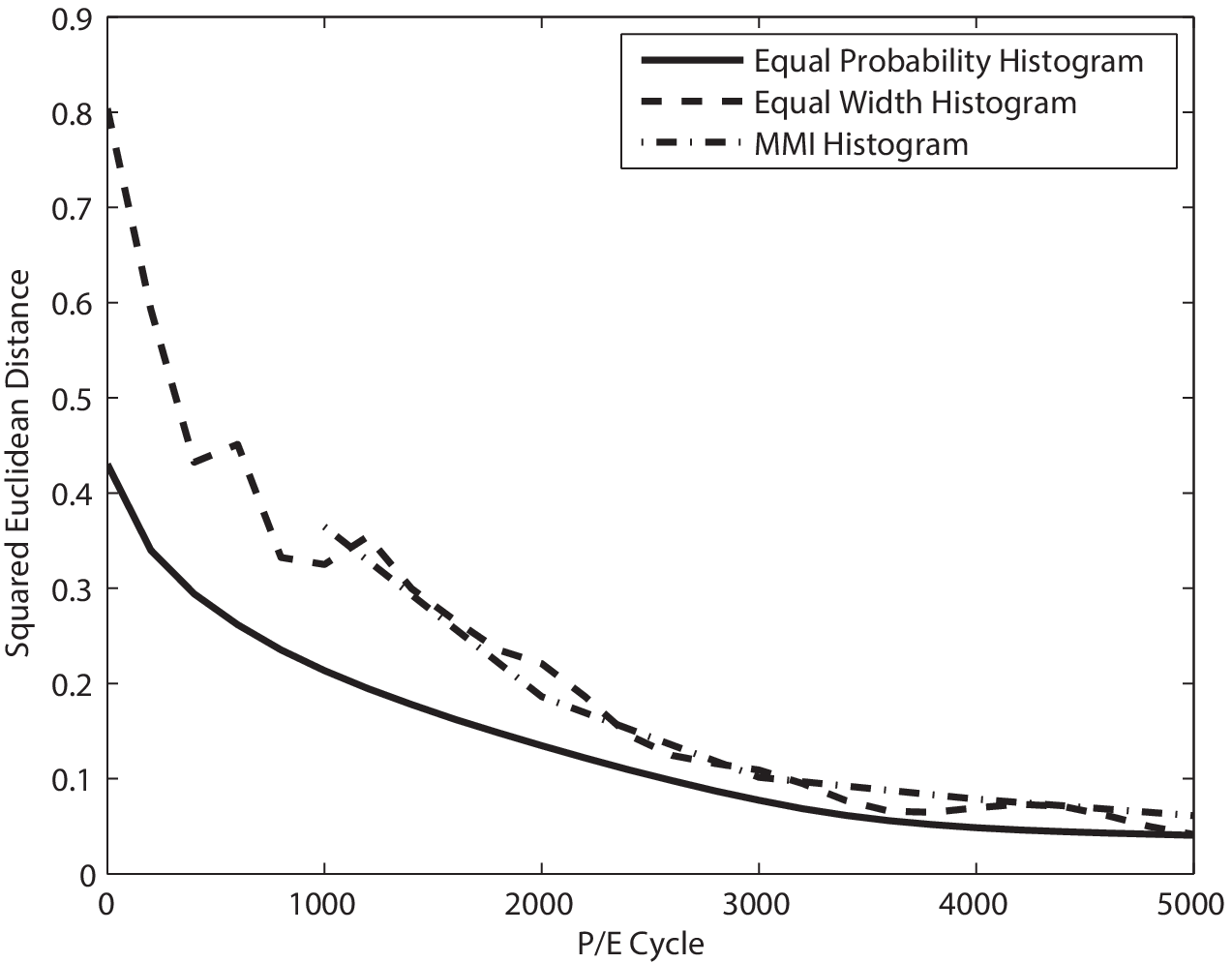}
\caption{Squared Euclidean distance between the channel distributions and corresponding histograms (10 bins).}
\label{Figure:SquaredEuclideanDistance_PDF&HistogramDensity}
\vspace{2ex}
\centering
\includegraphics[width=0.35\textwidth]{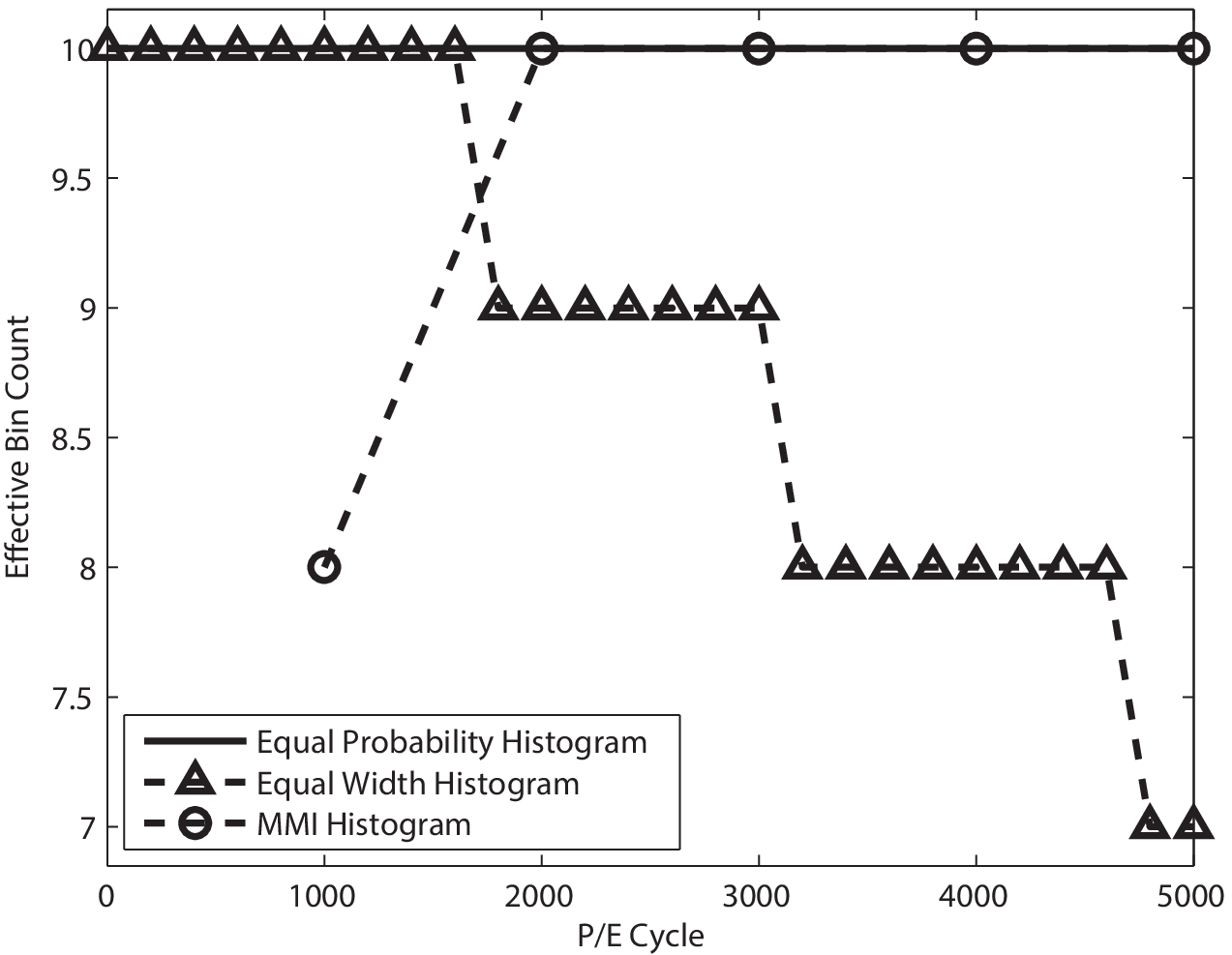}
\caption{Effective resolution of different histograms (10 bins).}
\label{Figure:EffectiveResolution}
\vspace{-2ex}
\end{figure}

Fig. \ref{Figure:SquaredEuclideanDistance_PDF&HistogramDensity} compares the metric $D_{E^2}$ generated from nine reads (ten bins) and the original channel distribution for the three bin-placement paradigms. The parameters provided in \cite{chen_increasing_2014} are used to generate evolving channel distributions as a function of the number of P/E cycles.  The equal-probability bin placement strategy provides a lower $D_{E^2}$ metric, and hence a better approximation to the original channel, than the other two strategies over a large span of P/E cycling conditions.  The performance difference is especially significant when the device condition is new. This behavior for all three paradigms is also seen when using histograms with 7 bins. As the number of bins grows, the performance difference becomes smaller, but we seek good performance with the fewest possible bins.  

In addition to the $D_{E^2}$ metric, effective resolution is used as a metric for the effectiveness of a bin placement strategy.  Because the value of each bin is always greater than or equal to zero, two adjacent zero-height bins can be combined as one bin. Thus, a word-line voltage at the boundary of two zero-count bins is a wasted read.  Although histogram bin probabilities derived from the integration will be nonzero in every interval, real measurements of a finite number of cells will produce zero-count bins. 

To compute the effective resolution, combine adjacent zero-count bins into one effective bin and count the number of resulting bins.  Fig.~\ref{Figure:EffectiveResolution} shows the effective resolution as a function of the number of P/E cycles for the three bin-placement paradigms.   Adjacent bins with induced probability less than $10^{-4}$ are combined. The equal-probability bin-placement paradigm has full resolution throughout the entire P/E cycling process, while the other paradigms lose resolution in some P/E cycling conditions. This suggests that the equal-probability bin-placement paradigm has a good tracking capability over the whole Flash lifetime.

Because it has superior performance both in terms of  $D_{E^2}$ and effective resolution, the equal-probability bin-placement paradigm is used in the channel parameter estimation discussed in the remainder of this paper.

\section{Channel Parameter Estimation}
\label{Section:ChannelParemeterEstimation} 

\subsection{Cost Function}
\label{Section:ChannelParemeterEstimation:HistogramCostFunctionAndJacobianMatrix}

From the discussion in Section \ref{Section:FlashMemoryChannelModel}, the channel parameter vector should be $[\lambda,\sigma_p,\sigma_e,\sigma_r,\mu_r]$.
However, both $\sigma_r$ and $\mu_r$ are intended-threshold-level dependent. As a result, $\mtx{\alpha}=[\lambda,\sigma_p,\sigma_e,\gamma_{\sigma_r},\gamma_{\mu_r}]$ is used in the following discussion as the level independent channel parameter vector to be estimated.  Level-independent channel parameters are preferred for supporting the DVA algorithm, which is a target application for HBCE.

Define $[q_0,q_1,\dots,q_M]$ as the boundaries of bins where $q_0=-\infty$, $q_M=\infty$, and $M$ is the number of bins. The number of cells in each bin can be estimated as
\begin{equation}
\label{Equation:BinEstimation}
\hat{N}_{bin,i}=\sum_{k=1}^{L}N_kP(q_i<y<q_{i+1}|x_k) \, ,
\end{equation}
where \mbox{$P(q_i<y<q_{i+1}|x_k)=\int_{q_i}^{q_{i+1}}f_{Y|X}(y|x_k) \,dy$} denotes the probability of a measured threshold falling in the $i$th bin when the intended threshold is $x_k$. $L$ is the number of intended threshold levels, and $N_k$ is the number of cells in each level determined by the stored data.

The cost function is defined as the normalized square Euclidean distance between the estimated histogram and the reference histogram 
\begin{equation}
C_M=\sum_{i=0}^{M-1}\left(\frac{N_{bin,i}-\hat{N}_{bin,i}}{N}\right)^2\, ,
\end{equation}
where $N$ is the total number of cells measured, and $N_{bin,i}$ is the $i$th bin's cell count in the reference histogram.  The gradient of the cost function is defined as 
\begin{equation}
\nabla \mtx{C}_M(\mtx{\alpha}) = 2 \cdot (\mtx{J}_{\mtx{G}_M}(\mtx{\alpha}))^T \cdot \mtx{G}_M(\mtx{\alpha}) \, ,
\end{equation}
where $\mtx{J}_{\mtx{G}_M}(\mtx{\alpha})$ is the Jacobian matrix of the difference vector between the estimated histogram and the reference histogram.

\subsection{Least Squares Algorithms}
\label{Section:ChannelParemeterEstimation:LeastSquaresAlgorithms}
Least squares algorithms have been widely used to fit a parameterized model to a data set. Three algorithms are examined in the following discussion.

\subsubsection{Gradient Descent (GD)}
\label{Section:ChannelParemeterEstimation:LeastSquareAlgorithms:GradientDescent}
GD minimizes the cost function by iteratively refining initial estimation of the parameters iteratively based on a linear model. In each iteration, the estimation is renewed by a step vector following the gradient of the cost function.

\begin{algorithm}
\caption{Gradient Descent Algorithm}\label{Algorithm:GradientDescent}
\begin{algorithmic}[1]
\State Initialize step size $\beta$ and $\mtx{\alpha}=\mtx{\alpha}^{(0)}$
\While{$\|\mtx{\alpha}^{(k+1)}-\mtx{\alpha}^{(k)}\|>\eta$ and $k < MaxIteration$}
  \State Compute $\mtx{J}_{\mtx{G}_M}(\mtx{\alpha}^{(k)}) $ and $\mtx{G}_M(\mtx{\alpha}^{(k)})$
  \State Compute $\nabla C_M(\mtx{\alpha}^{(k)})=2 \cdot (\mtx{J}_{\mtx{G}_M}(\mtx{\alpha}^{(k)})^T \cdot \mtx{G}_M(\mtx{\alpha}^{(k)})$
  \State $\mtx{\alpha}^{(k+1)}=\mtx{\alpha}^{(k)}-\beta \cdot \nabla \mtx{C}_M(\mtx{\alpha}^{(k)})$
  \State $k = k+1$
\EndWhile
\end{algorithmic}
\end{algorithm}

\subsubsection{Gauss-Newton (GN)}
\label{Section:ChannelParemeterEstimation:LeastSquareAlgorithms:GaussNewtonAlgorithm}
A quadratic model is employed to provide more accurate approximation of the cost function. The iterative relation can be represented as 
\begin{equation}
\mtx{\alpha}^{(k+1)}=\mtx{\alpha}^{(k)}-(\mtx{J}^T\mtx{J})^{-1}\mtx{J}^T\mtx{G} = \mtx{J}^+\mtx{G} \, ,
\end{equation}
where $\mtx{J}^+$ is the pseudo-inverse of $\mtx{J}$. Gauss-Newton Algorithm can then be formulated as follows:

\begin{algorithm}
\caption{Gauss-Newton Algorithm}\label{Algorithm:GaussNewtonAlgorithm}
\begin{algorithmic}[1]
\State Initialize $\mtx{\alpha}=\mtx{\alpha}^{(0)}$
\While{$\|\mtx{\alpha}^{(k+1)}-\mtx{\alpha}^{(k)}\|>\eta$ and $k < MaxIteration$}
  \State Compute $\mtx{J}_{\mtx{G}_M}(\mtx{\alpha}^{(k)})$ and $\mtx{G}_M(\mtx{\alpha}^{(k)})$
  \State $\mtx{\alpha}^{(k+1)}=\mtx{\alpha}^{(k)}-( \mtx{J}_{\mtx{G}_M}(\mtx{\alpha}^{(k)}))^+ \cdot \mtx{G}_M(\mtx{\alpha}^{(k)})$
  \State $k = k+1$
\EndWhile
\end{algorithmic}
\end{algorithm}

\subsubsection{Levenberg-Marquardt  (LM) \cite{marquardt_algorithm_1963}}
\label{Section:ChannelParemeterEstimation:LeastSquareAlgorithms:LevenbergMarquardtAlgorithm}
By combining GD and GN, LM possesses the advantages of both algorithms. The update vector $\mtx{\delta}_{\mtx{\alpha}}$ is calculated by solving $(\mtx{J}^T\mtx{J}+\beta\cdot diag((\mtx{J}_{\mtx{G}_M})^T\mtx{J}_{\mtx{G}_M}))\mtx{\delta}_{\mtx{\alpha}} = \mtx{J}^T\mtx{G}$ where $\beta$ acts as a weight to combine the two algorithms.
\begin{algorithm}
\caption{Levenberg-Marquardt Algorithm}\label{Algorithm:LevenbergMarquardtAlgorithm}
\begin{algorithmic}[1]
\State Initialize $\beta, v, \mtx{\alpha}=\mtx{\alpha}^{(0)}$ and $UpdateFlag=1$
\While{$\|\mtx{\alpha}^{(k+1)}-\mtx{\alpha}^{(k)}\|>\eta$ and $k < MaxIteration$}
	\If {$UpdateFlag = 1$}
		\State Compute $\mtx{J}_{\mtx{G}_M}(\mtx{\alpha}^{(k)})$ and $\mtx{G}_M(\mtx{\alpha}^{(k)})$
	\EndIf
	\State Solve $((\mtx{J}_{\mtx{G}_M})^T\mtx{J}_{\mtx{G}_M}+\beta\cdot diag((\mtx{J}_{\mtx{G}_M})^T\mtx{J}_{\mtx{G}_M}))\mtx{\delta}_{\mtx{\alpha}}=(\mtx{J}_{\mtx{G}_M})^T\mtx{G}_M$
	\State Compute $\mtx{J}_{\mtx{G}_M}(\mtx{\alpha}^{(k)})$ and $G_M(\mtx{\alpha}^{(k)})$
	\State $\mtx{\alpha}_{temporary}=\mtx{\alpha}-\mtx{\delta}_{\mtx{\alpha}}$
	\If {$\sum (err(\mtx{\alpha}))^2 > \sum (err(\mtx{\alpha}_{temporary}))^2$}
		\State $UpdateFlag=1$
		\State $\beta=\beta\cdot v$
		\State $\mtx{\alpha} = \mtx{\alpha}_{temporary}$
	\Else
		\State $UpdateFlag=0$
		\State $\beta=\beta / v$
	\EndIf
	\State $k = k+1$
\EndWhile
\end{algorithmic}
\end{algorithm}

\section{Simulation Results}
\label{Section:ExperimentalResults}

\begin{figure}
\centering
\includegraphics[width=0.35\textwidth]{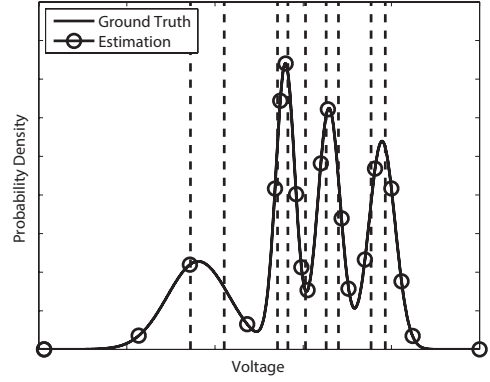}
\caption {Channel estimation result at 3000 P/E using Levenberg-Marquardt algorithm and 10-bin equal-probability histogram. (ground truth parameter vector is [0.0099,0.3500,0.0500,0.0617,-0.5882], estimation error vector is $10^{-4}\times$[0.0101 0.0214 -0.1774 0.0405 -0.0044])}
\label{Figure:ChannelEstimationExample}
\vspace{-2ex}
\end{figure}

To verify the effectiveness of the algorithms described in Section \ref{Section:ChannelParemeterEstimation} and determine the number of reads needed for equal-probability bin-placement paradigm, channel distributions generated with parameters from \cite{chen_increasing_2014} are used in our simulations. The retention time is set to be one year.  P/E cycling conditions from 0 to 4000 P/E are sampled every 300 P/E. The initial conditions for the three algorithms are the same [0.007,0.1,0.4,0.04,-0.4]. Figure \ref{Figure:ChannelEstimationExample} illustrates one of the estimation results of the 14 P/E cycling conditions.

\begin{table}[b]
\vspace{-4ex}
\renewcommand{\arraystretch}{1.3}
\caption{Converge counts of least square algorithms (over 14 cases).}
\label{Table:ConvergenceCount}
\centering
\begin{threeparttable}
\begin{tabular}{c|c|c|c}
\hline
No. of Reads & GD & GN & LM \\

\hline
6 & 0 & 1 & 12\\
\hline
9 & 0 & 3 & 13\\
\hline
12 & 0 & 4 & 11\\
\hline
\end{tabular}
\end{threeparttable}
\end{table}

\begin{figure}
\centering
\includegraphics[width=0.35\textwidth]{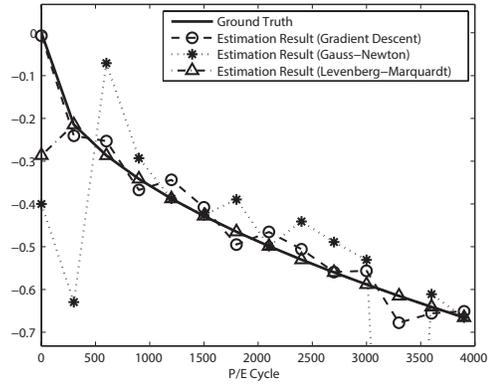}
\caption{Estimation result versus ground truth for $\gamma_{\mu_r}$ using 10-bin equal-probability histogram.}
\label{Figure:GammaMu_EstimationVSGroundTruth}
{\vspace{-3ex}}
\end{figure}

Figure \ref{Figure:GammaMu_EstimationVSGroundTruth} compares the estimating results of $\gamma_{\mu_r}$ with the ground truth, where the estimation algorithms employ 10-bin equal-probability histograms as input.  The LM algorithm performs significantly better than GD and GN in terms of both the estimation accuracy and the ability to adapt to different channel conditions. 

This behavior is further demonstrated in Table \ref{Table:ConvergenceCount}, which shows the convergence counts of the three algorithms over the 14 sample conditions. Every estimated parameter needs to be within $\pm$1\% of the ground truth parameter to qualify the estimated parameter vector as a converged result. GD fails in all simulation cases while reaching the maximum allowed number of iterations in every case. GN provides good results in certain cases with very few iterations. LM provides high estimation accuracy over different channel conditions, with some failures when the channel conditions are very good. Note that estimation accuracy of $\gamma_{\sigma_r}$ is usually higher than the other parameters. An intuitive explanation for this is that channel distribution mean shifts can be easily identified by even the histogram itself.

\begin{figure}
\centering
\includegraphics[width=0.40\textwidth]{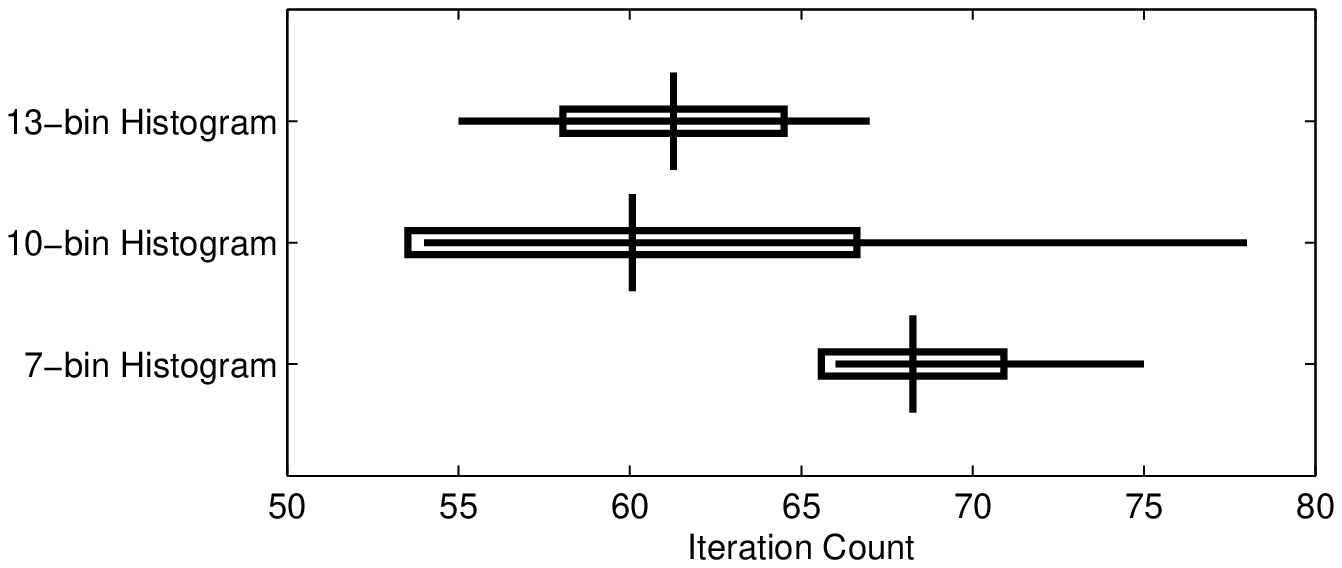}

{\tiny Horizontal Line Segment: Range of Iteration Count \\ Vertical Line Segment: Mean of Iteration Count\\ Rectangle: Standard Deviation of Iteration Count\par}
\caption{Iteration count statistics using Levenberg-Marquardt algorithm}
\label{Figure:LMIterationCount}
\centering
\includegraphics[width=0.45\textwidth]{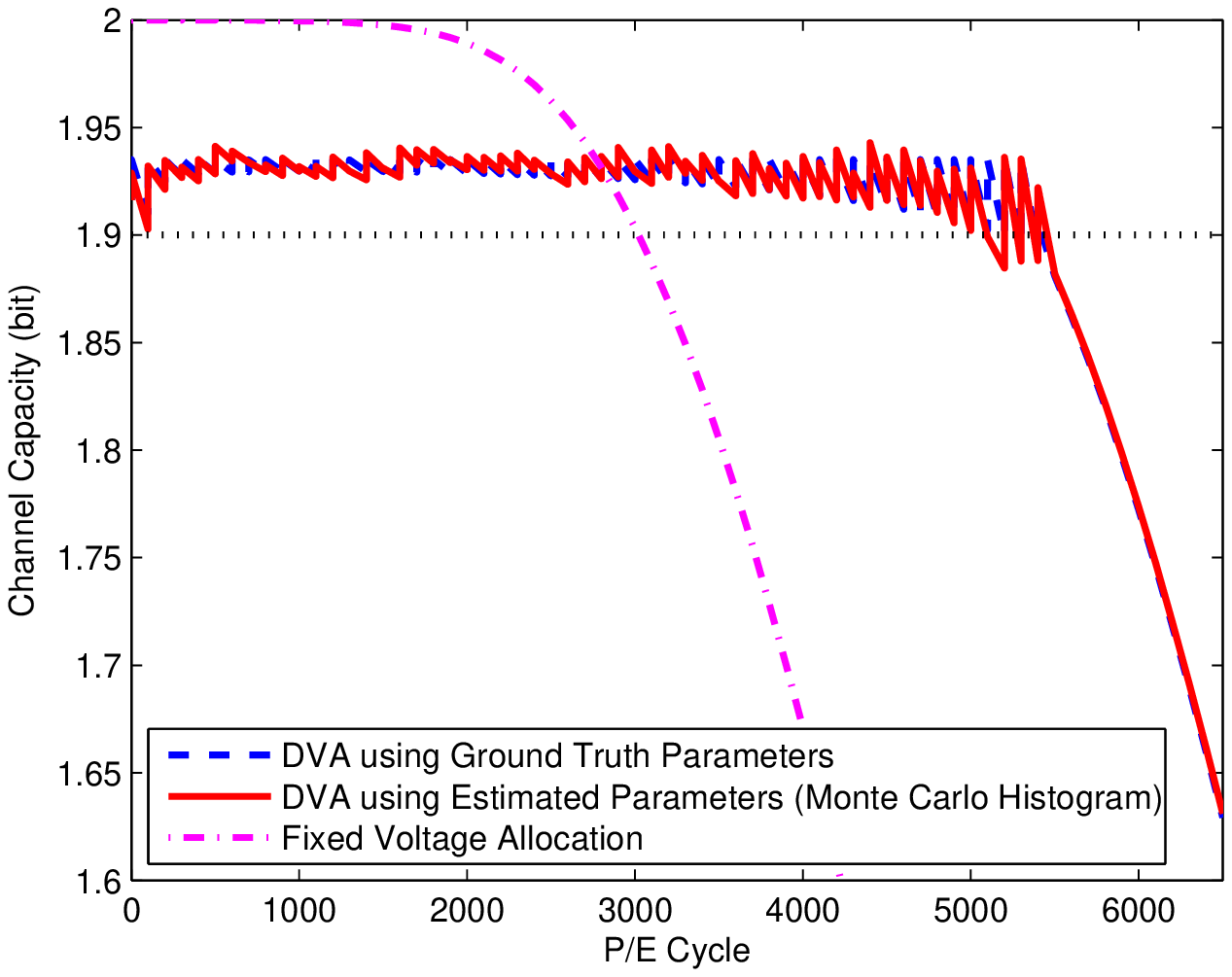}
\caption{Dynamic Voltage Allocation simulation result.}
\label{Figure:DVA_MonteCarlo}
\vspace{-2ex}
\end{figure}

Figure \ref{Figure:LMIterationCount} depicts key statistical metrics about the number of iterations when employing LM with histograms that differ in resolution over the 14 cases. The horizontal line segments represent the ranges of iteration counts, the vertical line segments indicate the mean values, and the rectangles show the standard deviations. 10-bin (9 reads) histograms reduce the number of iterations needed with respect to the results using 7-bin (6 reads) histograms. 13-bin (12 reads) histograms do not provide significant reduction in iteration counts, but do increase computational cost in each iteration. Overall, the 10-bin histograms provide a balance between iteration counts and the cost of each iteration.

\section{Conclusion}
\label{Section:Conclusion}
By combining measured histograms with least squares algorithms, we introduce a framework to estimate evolving Flash memory channel. Our analysis and simulation results show that good estimation accuracy and speed can be achieved by using Levenberg-Marquardt algorithm with 10-bin equal-probability histograms. With this framework, Flash channel estimation can be implemented with limited measurements. This enables further utilization of channel characteristics in future Flash solutions. We have successfully used the LM algorithm with a 10-bin histogram for a Dynamic Voltage Allocation \cite{chen_increasing_2014} algorithm. As shown in Figure \ref{Figure:DVA_MonteCarlo}, the performance using histogram generated by Monte Carlo histogram is not distinguishable from perfect knowledge of the channel.

\bibliographystyle{IEEEtran}
\bibliography{Flash_Model_Estimate_DVA_ICC}

% Generated by IEEEtran.bst, version: 1.13 (2008/09/30)
\begin{thebibliography}{10}
\providecommand{\url}[1]{#1}
\csname url@samestyle\endcsname
\providecommand{\newblock}{\relax}
\providecommand{\bibinfo}[2]{#2}
\providecommand{\BIBentrySTDinterwordspacing}{\spaceskip=0pt\relax}
\providecommand{\BIBentryALTinterwordstretchfactor}{4}
\providecommand{\BIBentryALTinterwordspacing}{\spaceskip=\fontdimen2\font plus
\BIBentryALTinterwordstretchfactor\fontdimen3\font minus
  \fontdimen4\font\relax}
\providecommand{\BIBforeignlanguage}[2]{{%
\expandafter\ifx\csname l@#1\endcsname\relax
\typeout{** WARNING: IEEEtran.bst: No hyphenation pattern has been}%
\typeout{** loaded for the language `#1'. Using the pattern for}%
\typeout{** the default language instead.}%
\else
\language=\csname l@#1\endcsname
\fi
#2}}
\providecommand{\BIBdecl}{\relax}
\BIBdecl

\bibitem{maeda_error_2009}
Y.~Maeda and H.~Kaneko, ``Error control coding for multilevel cell flash
  memories using nonbinary low-density parity-check codes,'' in \emph{Defect
  and Fault Tolerance in {VLSI} Systems, 2009. {DFT} '09. 24th {IEEE}
  International Symposium on}, Oct. 2009, pp. 367--375.

\bibitem{klove_systematic_2011}
T.~Klove, B.~Bose, and N.~Elarief, ``Systematic, single limited magnitude error
  correcting codes for flash memories,'' \emph{Information Theory, {IEEE}
  Transactions on}, vol.~57, no.~7, pp. 4477--4487, Jul. 2011.

\bibitem{chen_increasing_2014}
T.-Y. Chen, A.~R. Williamson, and R.~D. Wesel, ``Increasing flash memory
  lifetime by dynamic voltage allocation for constant mutual information,'' in
  \emph{Information Theory and Applications Workshop ({ITA}), 2014}, Feb. 2014,
  pp. 1--5.

\bibitem{sala_dynamic_2013}
F.~Sala, R.~Gabrys, and L.~Dolecek, ``Dynamic threshold schemes for multi-level
  non-volatile memories,'' \emph{Communications, {IEEE} Transactions on},
  vol.~61, no.~7, pp. 2624--2634, Jul. 2013.

\bibitem{bez_introduction_2003}
R.~Bez, E.~Camerlenghi, A.~Modelli, and A.~Visconti, ``Introduction to flash
  memory,'' \emph{Proceedings of the {IEEE}}, vol.~91, no.~4, pp. 489--502,
  Apr. 2003.

\bibitem{lee_estimation_2013}
D.-h. Lee and W.~Sung, ``Estimation of {NAND} flash memory threshold voltage
  distribution for optimum soft-decision error correction,'' \emph{Signal
  Processing, {IEEE} Transactions on}, vol.~61, no.~2, pp. 440--449, Jan. 2013.

\bibitem{lee_data_2003}
J.-D. Lee, J.-H. Choi, D.~Park, and K.~Kim, ``Data retention characteristics of
  sub-100 nm {NAND} flash memory cells,'' \emph{Electron Device Letters,
  {IEEE}}, vol.~24, no.~12, pp. 748--750, Dec. 2003.

\bibitem{monzio_compagnoni_random_2009}
C.~Monzio~Compagnoni, M.~Ghidotti, A.~Lacaita, A.~Spinelli, and A.~Visconti,
  ``Random telegraph noise effect on the programmed threshold-voltage
  distribution of flash memories,'' \emph{Electron Device Letters, {IEEE}},
  vol.~30, no.~9, pp. 984--986, Sep. 2009.

\bibitem{mielke_recovery_2006}
N.~Mielke, H.~Belgal, A.~Fazio, Q.~Meng, and N.~Righos, ``Recovery effects in
  the distributed cycling of flash memories,'' in \emph{Reliability Physics
  Symposium Proceedings, 2006. 44th Annual., {IEEE} International}, Mar. 2006,
  pp. 29--35.

\bibitem{cernea_34_2009}
R.-A. Cernea, L.~Pham, F.~Moogat, S.~Chan, B.~Le, Y.~Li, S.~Tsao, T.-Y. Tseng,
  K.~Nguyen, J.~Li, J.~Hu, J.~H. Yuh, C.~Hsu, F.~Zhang, T.~Kamei, H.~Nasu,
  P.~Kliza, K.~Htoo, J.~Lutze, Y.~Dong, M.~Higashitani, J.~Yang, H.-S. Lin,
  V.~Sakhamuri, A.~Li, F.~Pan, S.~Yadala, S.~Taigor, K.~Pradhan, J.~Lan,
  J.~Chan, T.~Abe, Y.~Fukuda, H.~Mukai, K.~Kawakami, C.~Liang, T.~Ip, S.-F.
  Chang, J.~Lakshmipathi, S.~Huynh, D.~Pantelakis, M.~Mofidi, and K.~Quader,
  ``A 34 {MB}/s {MLC} write throughput 16 gb {NAND} with all bit line
  architecture on 56 nm technology,'' \emph{Solid-State Circuits, {IEEE}
  Journal of}, vol.~44, no.~1, pp. 186--194, Jan. 2009.

\bibitem{cai_program_2013}
Y.~Cai, O.~Mutlu, E.~Haratsch, and K.~Mai, ``Program interference in {MLC}
  {NAND} flash memory: Characterization, modeling, and mitigation,'' in
  \emph{Computer Design ({ICCD}), 2013 {IEEE} 31st International Conference
  on}, Oct. 2013, pp. 123--130.

\bibitem{dong_using_2013}
G.~Dong, Y.~Pan, and T.~Zhang, ``Using lifetime-aware progressive programming
  to improve {SLC} {NAND} flash memory write endurance,'' \emph{Very Large
  Scale Integration ({VLSI}) Systems, {IEEE} Transactions on}, vol.~{PP},
  no.~99, pp. 1--1, 2013.

\bibitem{takeuchi_double-level-vth_1996}
K.~Takeuchi, T.~Tanaka, and H.~Nakamura, ``A double-level-vth select gate array
  architecture for multilevel {NAND} flash memories,'' \emph{Solid-State
  Circuits, {IEEE} Journal of}, vol.~31, no.~4, pp. 602--609, Apr. 1996.

\bibitem{compagnoni_first_2007}
C.~Compagnoni, A.~Spinelli, R.~Gusmeroli, A.~Lacaita, S.~Beltrami, A.~Ghetti,
  and A.~Visconti, ``First evidence for injection statistics accuracy
  limitations in {NAND} flash constant-current fowler-nordheim programming,''
  in \emph{Electron Devices Meeting, 2007. {IEDM} 2007. {IEEE} International},
  Dec. 2007, pp. 165--168.

\bibitem{wang_soft_2011}
J.~Wang, T.~Courtade, H.~Shankar, and R.~Wesel, ``Soft information for {LDPC}
  decoding in flash: Mutual-information optimized quantization,'' in
  \emph{Global Telecommunications Conference ({GLOBECOM} 2011), 2011 {IEEE}},
  Dec. 2011, pp. 1--6.

\bibitem{wang_enhanced_2014}
J.~Wang, K.~Vakilinia, T.-Y. Chen, T.~Courtade, G.~Dong, T.~Zhang, H.~Shankar,
  and R.~Wesel, ``Enhanced precision through multiple reads for {LDPC} decoding
  in flash memories,'' \emph{Selected Areas in Communications, {IEEE} Journal
  on}, vol.~32, no.~5, pp. 880--891, May 2014.

\bibitem{marquardt_algorithm_1963}
\BIBentryALTinterwordspacing
D.~W. Marquardt, ``\BIBforeignlanguage{English}{An algorithm for least-squares
  estimation of nonlinear parameters},''
  \emph{\BIBforeignlanguage{English}{Journal of the Society for Industrial and
  Applied Mathematics}}, vol.~11, no.~2, pp. pp. 431--441, 1963. [Online].
  Available: \url{http://www.jstor.org/stable/2098941}
\BIBentrySTDinterwordspacing

\end{thebibliography}

\end{document}